\documentclass[12pt]{article}
\usepackage{amssymb}
\usepackage{amsmath}
\newcommand {\ov} {\overline}
\newcommand{\ba}{\begin{eqnarray}}
\newcommand{\ea}{\end{eqnarray}}
\newcommand{\Ker}{\mathrm{Ker}}
\newcommand{\Coker}{\mathrm{Coker}}
\newsavebox{\uuunit}
\sbox{\uuunit}
    {\setlength{\unitlength}{0.825em}
     \begin{picture}(0.6,0.7)
        \thinlines
        \put(0,0){\line(1,0){0.5}}
        \put(0.15,0){\line(0,1){0.7}}
        \put(0.35,0){\line(0,1){0.8}}
       \multiput(0.3,0.8)(-0.04,-0.02){12}{\rule{0.5pt}{0.5pt}}
     \end {picture}}

\csname @addtoreset\endcsname{equation}{section}

\newcommand{\IC}{\mathbb{C}}
\newcommand{\IZ}{\mathbb{Z}}
\newcommand{\IP}{\mathbb{P}}
\def\Dslash{\not{\hbox{\kern-4pt $D$}}}
\def\dslash{\not{\hbox{\kern-2pt $\del$}}}
\begin{document}
\begin{titlepage}
\begin{flushright}
SU-ITP-05/10\\
SLAC-PUB-11074\\
hep-th/0503138
\end{flushright}
\vspace{.5cm}
\begin{center}
\baselineskip=16pt
{\bf \Large Counting fermionic zero modes on M5 with fluxes
}\\

\

{\large  Renata Kallosh,$^1$ Amir-Kian Kashani-Poor,$^{1,2}$ and Alessandro Tomasiello$^1$
 } \\

\

{\small 
$^1$ Department of Physics, Stanford University,
Stanford, CA 94305-4060, USA.\\ \vspace{6pt}
$^2$ SLAC, Stanford University, Stanford, CA 94305-4060, USA.
 }
\end{center}

\

\

\begin{center}
{\bf Abstract}
\end{center}
{\small   We study the Dirac equation on an M5 brane wrapped on a divisor in a Calabi--Yau fourfold in the presence of background flux. We reduce the 
computation of the normal bundle $U(1)$ anomaly to counting the solutions of a finite--dimensional linear system on cohomology. This system depends on the choice of flux. In an example, we find
that the presence of flux changes the anomaly and allows instanton corrections to the superpotential which would otherwise be absent. 
}
\vspace{2mm} \vfill \hrule width 3.cm
{\footnotesize \noindent e-mails: 
kallosh@stanford.edu, kashani@slac.stanford.edu, tomasiel@stanford.edu }
\end{titlepage}

\section{Introduction}

Some of the most interesting recent developments in string theory are related to the discovery of a mechanism for stabilizing moduli of dS vacua. Let us focus our attention on type IIB string theory.
There are two types of moduli here which have to be stabilized. Complex structure moduli and the axion-dilaton can be stabilized by introducing fluxes (electric and magnetic 3-forms) in the 10d theory \cite{Giddings:2001yu}. The 4d description of these systems is provided by no-scale supergravity. From the point of view of 10d supergravity,  fluxes are classical fields; they correspond to gauge coupling constants in the effective 4d theory. The second group of moduli, so-called K\"ahler moduli, cannot be stabilized by fluxes (i.e. gauging in 4d).\footnote{Note that this statement is not true in IIA compactification when world sheet instantons are taken into account, see \cite{Kachru:2004jr}.}. For cosmological applications, these are the most dangerous runaway fields with very steep potentials. For a long time, they presented a main impediment to finding dS vacua describing the present acceleration of the universe, as well as a stringy implementation of slow-roll inflation describing the early evolution of the universe. A basic mechanism for stabilizing these moduli was proposed in \cite{Kachru:2003aw} (the KKLT mechanism). However, it is challenging to prove conclusively that all relevant moduli can be stabilized; for recent progress in this direction, see \cite{Denef:2005mm} and references therein. 

There are two known sources of non--perturbative corrections  to the superpotential in no--scale models,  capable of stabilizing K\"ahler moduli. The first type of non--perturbative corrections to the superpotential comes from Euclidean D3 branes wrapping 4--cycles in a Calabi-Yau manifold (related to M5 branes wrapping 6-cycles in a dual picture). The second source of corrections to the superpotential, those stemming from gaugino condensation, is perhaps better understood, but it cannot be used in some interesting cosmological  models, e.g. the D3/D7 brane inflation models \cite{Herdeiro:2001zb,Angelantonj:2003zx,Koyama:2003yc}.  A particularly valuable feature of these model is that there is a shift symmetry for the inflaton field described by the distance between a D3 brane and a D7 brane. This symmetry provides the slow-roll conditions for early universe inflation without the fine-tuning required in many other models of stringy inflation. Due to their merits, it is desirable to find a mechanism which can fix the moduli fields that are not lifted by gaugino condensation (those pertaining to hypermultiplets) in these models. The investigation of this question was our main motivation for attempting to clarify the status of Euclidean D3 branes and M5 branes in compactifications with fluxes. 

In 1996, Witten
studied the conditions under which corrections to the superpotential can be generated by Euclidean D3 branes \cite{Witten:1996bn}, in the absence of background flux. His argument is based on the counting of zero modes of the Dirac operator on an M5 brane wrapped on a 6-cycle of a Calabi-Yau four-fold.
He finds that such corrections are possible only when the four-fold admits divisors $D$ of arithmetic
genus one, $\chi_D \equiv \sum (-1)^{n} h^{0,n}=1$. In the presence of such instantons, there is a correction to the
superpotential which at large volume yields a term
$
W_{\rm inst} = T(z_i) \exp(2\pi i \rho)\,.
$
Here, $T(z_i)$ is a complex structure dependent one-loop determinant,
and the leading exponential dependence comes from the action of a
Euclidean D3 brane wrapping a four-cycle.

Until recently, Witten's constraint $\chi_D =1$ was a guide for constructing all models in which K\"ahler moduli are stabilized via superpotentials generated by Euclidean instantons. However, investigation of D3/D7 inflation indicates that the condition $\chi_D =1$ may be too restrictive to allow the stabilization of all moduli in these cosmological models. 
Recently, it was argued  by  G\"orlich, Kachru, Tripathy and Trivedi \cite{Gorlich:2004qm} that in the presence of fluxes, a case not covered by Witten's analysis, there is a possibility of generating non-perturbative superpotentials even in models with branes wrapping divisors of arithmetic genus $\chi_D\geq 1$. Some aspects of instanton corrections in F-theory compactification with flux have also been studied recently in \cite{Robbins:2004hx}.

The goal of our paper is to perform a systematic investigation of this issue, and determine whether the condition $\chi_D=1$ for generating a superpotential may be relaxed in the presence of flux.  Our work is based  on the explicit Dirac action on the M5 brane in the presence of background fluxes obtained recently in \cite{Kallosh:2005yu}.
We will study how to count the number of zero modes of the flux modified Dirac operator.\footnote{Examples of such counting have recently been presented for fermions on D3 branes in \cite{Tripathy:2005hv}.} This will lead us to a generalization of the condition under which non-vanishing instanton corrections via Euclidean branes can be generated. In the absence of fluxes, our condition reduces to Witten's $\chi_D =1$ condition. 

Our  generalization of Witten's analysis to the presence of fluxes clarifies the necessary conditions for the existence of instanton corrections to the superpotential. This may play a crucial role in the  cosmological models in which instanton corrections provide the only known mechanism for fixing runaway moduli.

\section{Dirac equations on  M5 with background fluxes}

We will consider here only the simplest modification of the Dirac equation on M5 branes due to background fluxes; in particular, we ignore the  contribution of the 3-form on the worldvolume of the M5 brane.\footnote{The terms with the  3-form on the worldvolume of the M5 Dirac action depending on a background flux are known \cite{Kallosh:2005yu}, taking them into account requires a  separate investigation which we leave for future work.}  In this approximation, the Dirac part of the action is
\begin{equation}
L^{M5}_f={1\over 2}\,\theta\,\left[{\tilde
\gamma}^a \nabla_a\, +{1\over{24}}\,\left(\gamma^{ijk}\,
\tilde\gamma^{a}\,\,F_{aijk}-\gamma^i\,
\tilde\gamma^{bca}\,F_{abci}\right)\right]\,\theta\,.
\end{equation}
The Dirac equation on M5 reduces to 
\begin{equation}\label{wfe-h}
{\tilde \gamma}^{a\,\alpha\beta} \,\nabla_a\, \theta_{\beta q}
+{1\over{24}}\,[\,(\gamma^{ijk})_{q p} \,
(\tilde\gamma^{a})^{\alpha \beta}\,F_{aijk}-\,(\gamma^i)_{q p}\,
(\tilde\gamma^{abc})^{\alpha \beta} \,F_{abci}] \theta_{\beta}{}^ p
=0\,.
\end{equation}
Here $a$ with $a=0,...,5$ are directions on the brane and $i$ with $i=1,...,5$ are directions normal to the brane.
On the M5 worldvolume there is a local supervielbein frame $e^{A}= (e^a, e^{\alpha q})$  with mostly positive signature of the metric  and its components form
the vector and spinor representations of the group $SO(1,5)\times SO(5)$. The index $\alpha$ belongs to a representation of $ Spin(1,5)$ and the index $q$ to the representation of $ Spin(5)$,  $ \alpha=1,...,4$ and  $ q=1,...,4$. Spinors in d=6 are $USp(4)$ symplectic Majorana-Weyl spinors.  Therefore $\theta^{\alpha q}$ and $\theta^q_{\alpha}$ have opposite $Spin(1,5)$ chiralities. Spinors 
$\theta^{\alpha q}$ and $\theta^{\alpha}_q$ are related by the $USp(4)$ invariant tensor $C_{pq}$. When $Spin(5)$ is broken down to $Spin(3)\times Spin(2)$ the spinors with up and down position of the index have opposite chiralities in $Spin(2)$.
The gauge-fixing of $\kappa$-symmetry leaves only $\theta_{\beta q}$ and $\theta_{\beta}{}^ p$ spinors on the brane, i. e. only one type of chirality in $Spin(1,5)$ and both types of chirality on $Spin(2)$.
We refer to \cite{Kallosh:2005yu} and to the review paper \cite{Sorokin:1999jx} for details.

To convert this construction to Euclidean space we have to replace $\gamma^0$ from the group of $\gamma^a$ matrices on the worldvolume to $\gamma^0_E= \pm i \gamma^0$ and of course do the same in the target space of 11d supergravity. $SO(1,5)$ will become $SO(6)$, $Spin(1,5)$ will become $Spin(6)$ etc. and the signature of the metric will be all positive.

\subsection{U(1) symmetry in the presence of fluxes and F-chirality}

The action is invariant and equations of motion covariant  under the $SO(5)$ structure group which is the reflection of the symmetries  of 11-dimensional supergravity. The flux $F_{abci}$ transforms as a vector and $F_{ijka}$ as an antisymmetric tensor of the 3d rank. In the case of interest  the structure group $SO(5)$ is broken down to $SO(3)\times SO(2)$.

In Witten's analysis  \cite{Witten:1996bn}  of instanton generated superpotentials, the $SO(2)\sim U(1)$ structure group of the normal bundle plays the role of an R-charge for the 3 dimensional effective theory. Via a sequence of arguments, the R-charge of the superpotential of the 3d theory is related to the number of zero modes of the Dirac equation on the M5 brane (which is twisted by the spin bundle associated to the normal bundle; the spinors take values in $S_+\otimes(S'_+ \oplus S'_-)$, with $S_+$ the positive chirality spin bundle associated to the tangent bundle of the divisor $D$, and $S'_+$, $S'_-$ the spin bundles of respective chirality associated to the normal bundle of $D$), counted with sign according to their $U(1)$ charge. Witten demonstrates that this quantity is given by the arithmetic genus $\chi_D$ of the divisor the M5 brane is wrapping. We review his argument in section (\ref{sec:noflux}). This charge depends on the chirality of the zero modes in the normal direction. The $\IZ_2$ subgroup of $U(1)$ (rotation by $\pi$) is, in fact, generated by $\Gamma$, the `$\gamma_5$' of the normal direction. In the presence of flux, the solutions of the Dirac equation no longer have definite chirality (in the following, unless otherwise noted, chirality will always mean chirality in the normal direction), as $\Gamma$ no longer commutes with the Dirac operator. Schematically,
\ba
\Gamma ( \Dslash + \Gamma^i F_i ) &=& ( \Dslash - \Gamma^i F_i ) \Gamma   \,.
\ea
This is not surprising, since $F$, as a vector in the normal direction, also transforms under $U(1)$. The action of the $\IZ_2$ subgroup of $U(1)$ on the Dirac operator in the presence of flux is hence generated by $\Gamma$ with simultaneous rotation of $F$ by $\pi$, i.e. $F\rightarrow -F$. We will refer to this $\IZ_2$ symmetry as F-chirality. Let us stress that the invariance of the Dirac operator under F-chirality is no more than its invariance under (a $\IZ_2$ subgroup, for convenience) of the structure group of the normal bundle. Since F-chirality commutes with the Dirac operator, we can choose a basis of solutions of the Dirac equation with definite charge under F-chirality. To be specific, let's denote the solutions of the Dirac equation, which are generically going to be functions of $F$, by $\epsilon(F)$. We can choose a basis of solutions such that $\Gamma \epsilon(-F) = \pm \epsilon(F)$. We refer to the number of solutions to the Dirac equation in the presence of flux weighted by F-chirality as $\chi_D(F)$.

Note that we cannot use a homotopy argument built around deforming $F$ continuously to 0 to conclude that $\chi_D$ and $\chi_D(F)$ must coincide. We will return to this point at the end of section \ref{sec:noflux} after we have brought the Dirac equation into a more palatable form.

\section{Solving the spinor equations}
The goal of this section is to reduce the problem of determining the F-index of the Dirac operator in the presence of flux to counting the solutions of a linear system.
We begin in section \ref{sec:explicit} by rewriting the equations in a form amenable to  manipulation. 
In \ref{sec:noflux}, we recall how the solutions are counted in terms of
cohomology for the case without flux. In \ref{sec:separate} we turn
to the case with flux, but with the simplifying assumption that the
fluxless Dirac operator and the action of the flux on the spinor are separately zero. 
Finally, in \ref{sec:general}, we give the 
general solution in terms of a finite--dimensional linear system, in 
particular arguing for the number of solutions in the generic case;
in \ref{sec:counting}, we argue how these solutions should be counted
with sign, using F-chirality as discussed in the previous section.

\subsection{The equations made explicit in terms of forms}
\label{sec:explicit}

We will be considering fluxes of the type $F_{abci}$, in the notation introduced in the previous section. Moreover, requiring the preservation of $N=1$ supersymmetry (we use 4d terminology throughout this paper, i.e. $N=1 \rightarrow$ 4 supercharges) imposes that they are (2,2) and primitive \cite{Becker:1996gj}. 
Under these assumptions, we show  
that the Dirac equations on the M5 brane with 
fluxes are of the following form
\begin{equation}
 D \epsilon_+=0\, , \qquad D\epsilon_-=F \epsilon_+
\label{Diracwithfluxes}
\end{equation}
Here, the subscript on $\epsilon_{\pm}$ refers to the chirality in the normal directions (both $\epsilon_{\pm}$\ have positive chirality on the M5 worldvolume). 

Let us review the setup.
The M5 brane is wrapped on a 6-cycle inside the compactifying 4-fold. 
The 8 dimensions of the 4-fold include 
the directions tangent to the M5, $a,b,c, \ov a,  \ov b, \ov c$, and normal to it, $z, \ov z$. The spinors on the worldvolume are twisted by the spin bundle associated to the normal bundle. Following Witten \cite{Witten:1996bn}, we identify the bundle $S$ in which the spinors take values as
\ba
S &=& S_+ \otimes (N^{1/2} \oplus N^{-1/2}) \\
&=& (K^{1/2} \oplus  K^{1/2} \otimes \Omega^{0,2} ) \otimes (K^{1/2} \oplus K^{-1/2})\\
&=& {\cal O} \oplus \Omega^{0,2} \oplus K \oplus (\Omega^{0,2} \otimes K) \,.
\ea
We have here used the triviality of the canonical bundle of the ambient space and the adjunction formula to identify $N=K$.
 We now use standard arguments about spinors on 
K\"ahler manifolds to make the identification between spinors and forms explicit. To this end, we let the gamma matrices act as 
\begin{equation}
  \label{eq:clifford}
\Gamma^{\ov a} = dz^{\bar a}\wedge\ , \qquad 
\Gamma^a=g^{a\bar b} \iota_{\bar b}\ ,   
\end{equation}
where 
$\iota_{\ov b} dz^{\bar a_1} \wedge\ldots \wedge dz^{\bar a_k} = 
k\delta_{\ov b}{}^{[ \ov a_1} dz^{\ov a_2}\wedge \ldots \wedge 
dz^{\bar a_k]}$, and define a Clifford vacuum as a state
$|\Omega\rangle$ that satisfies
\begin{equation}
\Gamma^{ z}|\Omega\rangle=0\,, \qquad \Gamma^{ a}|\Omega\rangle=0\ .
\label{annih}
\end{equation}
Since the spinors on M5 have positive worldvolume chirality, our spinors will always have an even number of worldvolume $\gamma$-matrices. The chirality in the normal direction, which will play a crucial role in our investigations, is determined by whether we have an even or odd number of normal bundle $\gamma$-matrices.
Explicitly, the decomposition of the spinor is
\begin{equation}
  \epsilon_+= \phi |\Omega\rangle + \phi_{\ov a \ov b}\Gamma^{\ov a \ov b}|\Omega\rangle\ ;
\label{fermions+}
\end{equation}
\begin{equation}
  \epsilon_-= \phi_{\ov z} \Gamma^{\ov z}|\Omega\rangle + 
  \phi_{\ov z \ov a \ov b}\Gamma^{\ov z \ov a \ov b}|\Omega\rangle\ .
\label{fermions-}
\end{equation}
(\ref{fermions+}) and (\ref{fermions-}) are respectively sections of $\Omega^{0,ev}$ and
$\Omega^{0,ev}\otimes K$. Concretely, we can see the latter in the fact that
the forms in (\ref{fermions-}) have an index $_{\ov z}$. 

The Dirac equation with fluxes acting on anti-holomorphic forms reads
\begin{equation}
\Gamma^a \nabla_a \epsilon 
+ \Gamma^{\ov a}\nabla_{\ov a}\epsilon + F \epsilon=0\ ,
\label{dirac}
\end{equation}
where $F= F_{\ov a\ov z bc}\Gamma^{\ov a\ov z bc}+ 
F_{\ov b\ov c az}\Gamma^{\ov b\ov c az}$. Using the explicit Clifford 
representation (\ref{eq:clifford}), we get
\begin{equation}
[\partial_{\ov a}\phi+4 g^{\ov b c}\partial_c
\phi_{\ov b \ov a}]\Gamma^{\ov a}|\Omega\rangle=0
\label{dirac1}
\end{equation}
\begin{equation}
[ \partial^A_{\ov a}\phi_{\ov z} + 
4 g^{\ov b c} \partial^A_c\phi_{\ov b \ov a \ov z} 
+ 8F_{\ov a \ov z bc}\phi^{bc}]\Gamma^{\ov a \ov z}|\Omega\rangle =0
\label{dirac2}
\end{equation}
\begin{equation}
[\partial_{\ov a}\phi_{\ov b \ov c}  \Gamma^{\ov a \ov b \ov c }]
|\Omega\rangle =0
\label{dirac3}
\end{equation}
\begin{equation}
  \label{dirac4}
 [\partial^A_{\ov a} \phi_{\ov b \ov c \ov z} ]
\Gamma^{\ov a \ov b \ov c \ov z}
|\Omega\rangle \ .
\end{equation}

On forms which also have a $_{\ov z}$ index, namely forms in 
$\Omega^{0,ev}\otimes N$, we have a covariant derivative $\partial^A 
\equiv\partial +A$ rather than the straight derivative. 
$A$ is a connection on the line bundle $K$, which therefore acts by 
multiplication. It plays no explicit role 
in the following, as we will see. If one prefers, one can even make it 
 disappear by mapping these $N$--valued forms to ordinary ones as in 
(\ref{eq:map}).

Notice that $F \epsilon_-$ has turned out to be zero identically, as we had
promised in (\ref{Diracwithfluxes}). There are 
several ways of seeing this. One is to use anti--imaginary--self--duality and 
the chiral projector on the spinor. Another is to count the number of 
barred gamma matrices: using primitivity, we have 
$F_{\ov b\ov c az}\Gamma^{\ov b\ov c az}= F_{\ov b\ov c az}
\Gamma^{az}\Gamma^{\ov b\ov c}$ or also $F_{\ov b\ov c az}\Gamma^{\ov b\ov c} 
\Gamma^{az}$; we can then use one form or the other to show that it annilates 
both summands in (\ref{fermions+}). It is also instructive to use SU(3) group theory (even if the canonical line bundle $K$ is not 
trivial). 
$F_{\ov b \ov c d z}$, 
disregarding the $z$ index, is $(1,2)$ and primitive, which is a $6$. It acts
on the state $\Gamma^{\ov d \ov e}|\Omega\rangle$, which is a $3$. The 
alleged result, $F_{\ov b \ov c dz}\phi_{\ \ov z\ov a}^d$, 
would be a $(0,3)$ form, 
that is, a singlet. But there is no singlet in $6\otimes 3=
10\oplus 8$. (If $F$
had a non--primitive part, it would be a $\bar 3$, and there would be a 
singlet in $\bar 3\otimes 3 = 8 \oplus 1$.) Finally, the brute force way of obtaining this result is the following. Evaluating $F$ 
acting on $\phi_{\ov z \ov a \ov b}$ yields
\ba \label{thelongway}
F_{\ov b \ov c a z} \phi^{z a}{}_{\ov d} \,\Gamma^{\ov b\ov c \ov d} 
| \Omega \rangle \,.
\ea
Let us now consider frame indices so that the complex structure is just 
that of $\IC^4$. The primitivity condition on $F$ then implies
\ba
\sum_{\ov a} F_{\ov b \ov a z}{}^{\ov a} = 0  \, .
\ea
Writing out (\ref{thelongway}), we get a sum of six terms, the terms 
of which pair up to give zero. One such pair is 
\ba
F_{\ov 1 \ov 2 2 z} \phi^{z 2}{}_{\ov 3} + F_{\ov 3 \ov 1 3 z} 
\phi^{z 3}{}_{\ov 2} &=& F_{\ov 1 \ov 2 z}{}^{\ov 2} 
\phi^{z}{}_{\ov 2\ov 3} + F_{\ov 3 \ov 1 z}{}^{\ov 3} 
\phi^{z}{}_{ \ov 3 \ov 2} = 0  \,.
\ea

\subsection{Case without fluxes}
\label{sec:noflux}

In absence of fluxes, the equations (\ref{dirac1}) -- (\ref{dirac4}) reduce to 
\begin{equation}
  \label{eq:dirac0}
  \partial_{\ov a}\phi+4 g^{\ov b c}\partial_c
\phi_{\ov b \ov a}= 0\ ,\qquad \partial^A_{\ov a}\phi_{\ov z} + 
4 g^{\ov b c} \partial^A_c\phi_{\ov b \ov a \ov z} = 0 \ , \qquad
\partial^A_{[\ov a} \phi_{\ov b \ov c] \ov z}= 0  \ , \qquad
\partial_{[\ov a}\phi_{\ov b \ov c]}=0\ .
\end{equation}
We first review how to reduce these equations to the statement that the forms $\phi, \phi_{\ov z}, \phi_{\ov a \ov b}, \phi_{\ov a \ov b \ov z}$ are harmonic. Let us introduce the operator $dz^{\ov a} \partial_{\ov a}
\equiv \bar\partial$ and its adjoint $\bar\partial^\dagger$. 
If we now act on the equations (\ref{eq:dirac0}) with $\bar\partial +
\bar\partial^\dagger$, using $\bar\partial^2=0, (\bar\partial^\dagger)^2=0$,
we get separate equation for each form, such as 
$\bar\partial^\dagger\bar\partial \phi=0$. These equations together imply 
that the laplacian $\Delta\equiv (\bar\partial + \bar\partial^\dagger)^2=
\bar\partial \bar\partial^\dagger +\bar\partial^\dagger \bar\partial $
of all the forms vanishes. Hence they are harmonic. 

To count the number of solutions to these equations, weighted by chirality, we would like to dispose of $N$--valued forms by mapping 
them to ordinary ones. This is possible due to Serre duality at the level of forms (rather than cohomology) which yields
\ba
\Omega^{0,ev}\otimes K\cong \Omega^{0,odd} \,.  \label{serre}
\ea
 Concretely, 
let us look for example at $\phi_{\ov a \ov b \ov z}$. The fact that $N=K$ 
simply means that we can multiply it by $\Omega_{abcz}$, the holomorphic 
four--form in the ambient Calabi--Yau. Then we can apply the Hodge star. 
The result is the mapping
\begin{equation}
  \label{eq:map}
\phi_{\ov a \ov b \ov z} \mapsto \tilde\phi_c=\Omega_{abcz}\phi^{abz}\ ;
\qquad \phi_{\ov z} \mapsto \tilde\phi_{abc}= \Omega_{abcz}\phi^z \ .  
\end{equation}
We have obtained forms of type $(odd,0)$ rather than $(0,odd)$, which changes
nothing since the submanifold is K\"ahler. We have used the physicists' Hodge
star rather than the mathematicians' one, which would have had an extra
conjugation.

This mapping can be used to count forms in the case without flux. 
As we have seen at the beginning of this subsection, without fluxes we 
get that the forms which represent the spinors are actually harmonic. The 
even forms $\phi$, $\phi_{\ov a \ov b}$ are counted by $H^{0,0}$ and
$H^{0,2}$ respectively. These forms obviously have positive normal--bundle
chirality. Then we have $\phi_{\ov z}$ and $\phi_{\ov a \ov b \ov z}$, with
negative normal--bundle chirality. As we have just seen, these are mapped
to $H^{3,0}$ and $H^{1,0}$ respectively. The index thus results to be
\begin{equation}
  \chi_D= (-)^p h^{(0,p)}
\label{index}
\end{equation}
which is called the holomorphic characteristic, or arithmetic genus of the divisor $D$. 

The rest of this section is devoted to studying how this index changes in the presence of fluxes. (One cannot use any homotopy
arguments to argue that the index does not change, because the new term 
with $F$ connects chiralities in a way different from Dirac; so the total
operator acts on a different space than the one used to define the usual 
index.) 

Before proceeding with the analysis of our equations in the presence of flux, we want to return to the question of the homotopy equivalence of the Dirac operator in the absence/presence of flux.
In the usual case without fluxes, what we want to count is the 
difference between the number of solutions to the Dirac equation with positive chirality 
minus the number with negative chirality. Calling $D_\pm$ the restriction of 
$D$ to positive or negative chirality, this difference
is $\mathrm{dim}(\Ker(D_+))
-\mathrm{dim}(\Ker(D_-))$. This can then be reformulated as
$\mathrm{dim}(\Ker(D_+)) -\mathrm{dim}(\Coker(D_+))$, since $D_-=D_+^\dagger$. 
Here we are using the 
isomorphism (\ref{serre}) to reexpress $D_-: \Omega^{0,ev}\otimes K \to 
\Omega^{0,odd} \otimes K$ as $D_-:\Omega^{0,odd} \to \Omega^{0,ev}$.  
In the presence of flux however, the $F$ dependent piece of the Dirac 
operator maps $\Omega^{0,ev} \rightarrow \Omega^{0,ev}$. The introduction of the operators $(D+F)_+$ and $(D+F)_-$ is no longer appropriate, as the operator $D+F$ may have mixed chirality zero modes. This is a reformulation of the statement that chirality no longer commutes with the Dirac operator in the presence of flux. As we have argued above, the appropriate index in the presence of flux is the number of solutions with positive
F--chirality minus negative F--chirality. This index cannot be formulated as 
$\mathrm{dim}(\Ker(O)) -\mathrm{dim}(\Coker(O))$ for a suitable operator $O$. Hence, no homotopy argument applies.

\subsection{Separate solutions}
\label{sec:separate}

As a warm--up, let us consider solutions which separately satisfy $D (\epsilon_+ +\epsilon_-)=0$ and $F (\epsilon_+ + \epsilon_-)=0$ (we will call them ``separate solutions''). Separate solutions are hence solutions of the fluxless Dirac equations which satisfy the additional constraint $F \epsilon_+=0$. They hence clearly do not exhibit a functional dependence on $F$, and the notion of F-chirality therefore reduces to ordinary chirality (which, recall, we are taking to mean chirality in the normal direction).
To analyse the constraint $F \epsilon_+=0$, we extract the only piece with $F$ dependence from 
(\ref{dirac3}), which leads to
\begin{equation}
 F_{\ov a\ov z bc} \phi^{bc}=0\ .
\label{2'}
\end{equation}
This equation is algebraic. Except for special choice of flux, it will only have the trivial solution. 

\subsection{ Harmonic projectors, and the general case}
\label{sec:general} 

We will here present a general analysis of the equations (\ref{dirac1}--\ref{dirac4}). 

The equations read $D \epsilon_+=0, D\epsilon_-=F \epsilon_+$. 
As we have already observed, $F \epsilon_+ =0$ identically. Hence we can analyze the equation $D\epsilon_+=0$, that is, equations (\ref{dirac1}) and 
(\ref{dirac3}), just as in the fluxless case. As in
subsection \ref{sec:noflux}, we see that these two together 
imply that $\phi$ and $\phi_{\ov a \ov b}$ are harmonic.
To make progress with the remaining equation, $D\epsilon_-=F\epsilon_+$, we notice that equation (\ref{dirac2}) has the schematic from $\partial \phi_1 + \partial^\dagger \phi_3 + F \phi_2 =0$. 



We now
introduce the projector
${\cal H}$ onto harmonic forms, i.e. ${\cal H}(\omega)$
is a harmonic form (possibly zero) for any form $\omega$. 
This projector gives zero on any exact or co--exact form:
\begin{equation}
  \label{eq:ex}
  {\cal H} \bar\partial \omega=0\ ; \qquad
  {\cal H} \bar\partial^\dagger \omega=0\qquad \forall \omega\ :
\end{equation}
exact forms are trivial in cohomology, and co--exact forms 
are trivial in $\bar\partial^\dagger$ cohomology.

Let us now look at eqns. (\ref{dirac2}) and (\ref{dirac4}) (the connection 
$A$, as we have said, is not essential; we can either include
it in the definition of a slightly different laplacian $\Delta^A$, or 
choose to use (\ref{eq:map}) and express the equations in terms of odd forms
$\tilde\phi_a$, $\tilde\phi_{abc}$). Let us act on (\ref{dirac2}) with the 
zero--mode projector ${\cal H}$. Then, thanks to (\ref{eq:ex}), we obtain
\begin{equation}
  \label{eq:harm}
  {\cal H} (F_{\ov a \ov z bc}\phi^{bc}dz^{\ov a})=0\ .
\end{equation}
We can interpret this equation as a linear operator ${\cal H}F$ annihilating
the form $\phi_{\ov a \ov b}$. The space in which this operator ends is
$H^{0,1}(D, N)$, that is, again $H^{2,0}$, if we apply again the dualization 
procedure (\ref{eq:map}). So the equation (\ref{eq:harm}) gives $h^{2,0}$ 
equations for $h^{2,0}$ variables. 

Are these equations linearly independent? This is hard to determine without knowing 
more about $F$. We can find no obvious arguments 
(such as group theoretical ones) for a linear dependence. We conclude 
that, for a generic choice of $F$ (that is, in an open set in the space of 
all $F$'s) the system is of maximal rank, and hence (being square) admits no
solutions. This kills all of the $\phi_{\ov a \ov b}$. Let us stress once 
again that there will be situations, possibly interesting ones, in which
the rank will go down, and some non--zero solutions for $\phi_{\ov a \ov b}$
will be possible, all the way down to the fluxless case $F=0$, in which all of them
are solutions.


Having determined the constraint of $\phi_{\ov a \ov b}$, we can now find solutions for $\phi_{\ov z}$ and $\phi_{\ov a \ov b \ov z}$ when a $\phi_{\ov a \ov b}$ exists that satisfies this constraint.

Let us consider the projector $1-{\cal H}$ onto non-zero modes of the laplacian. Obviously, on the 
image of this projector the laplacian is invertible: every operator is 
invertible away from its kernel.  The inverse is the Green 
operator $G$. In formulae, $1-{\cal H}= \Delta G$. Or, 
\begin{equation}
  \label{eq:hdec}
  (1 -{\cal H})(\omega) = 
\bar\partial ( \bar\partial^\dagger G \omega)
+ \bar\partial^\dagger (\bar\partial G\omega)\ \qquad \forall \omega\ .
\end{equation}
Let us now take $\omega= F_{\ov a \ov z bc}\phi^{bc}dz^a$. 
Comparing this with (\ref{eq:harm}), we see that we can actually drop 
${\cal H}$ from the left--hand--side of (\ref{eq:hdec}). 
Looking at (\ref{dirac2}), we see that (\ref{eq:hdec}) gives us a solution:
\begin{equation}
  \label{eq:partsol}
  \phi^0_{\ov z}= g^{b \ov a} \partial_b 
\left(G F_{\ov a \ov z cd}\phi^{cd}\right)\ ;
 \qquad \phi^0_{\ov a \ov b\ov z}= \partial_{[\bar b} 
\left(G F_{\ov a] \ov z cd}\phi^{cd}\right)\ .
\end{equation}
The superscript $^0$ is to emphasize that this is a particular solution; not 
the most general. However, we know that once a particular solution to an 
equation is found, we can find the most general one by solving the 
corresponding homogeneous equation. In our case, that amounts to 
solving (\ref{dirac2}) with $F=0$ (since we want it to be homogeneous now)
and (\ref{dirac4}). These can again be shown (as above, by acting with 
$\bar\partial + \bar\partial^\dagger$) to have, as most general 
solution, $\phi_{\ov z}$ and $\phi_{\ov a \ov b\ov z}$ to be harmonic; 
there are, as we have seen using (\ref{eq:map}), respectively $h^{3,0}$ and
$h^{1,0}$ such forms. 

To summarize, the solution space of the Dirac equation, denoting an element of this space as
$\left(
\begin{matrix}
\phi ,
\phi_{\ov a \ov b},
\phi_{\ov z} ,
\phi_{\ov a \ov b \ov z} ,
\end{matrix}
\right)^T, $
is spanned by the vectors

\begin{equation}
  \label{eq:recap}
 \left\{ \begin{array}{c}\vspace{.2cm}
\left(
\begin{matrix}
\phi^{harm}\\
0\\
0 \\
0 \\
\end{matrix}
\right)  
\;, \qquad 
\left(
\begin{matrix}
0\\
0\\
\phi_{\ov z}^{harm} \\
0 \\
\end{matrix}
\right)  \;, \qquad
\left(
\begin{matrix}
0\\
0\\
0 \\
\phi^{harm}_{\ov a \ov b\ov z}\\
\end{matrix}
\right)
\;,
\qquad 
\left(
\begin{matrix}
0\\
\tilde{\phi}_{\ov a \ov b}\\
g^{b \ov a} \partial_b 
\left(G F_{\ov a \ov z cd}\tilde{\phi}^{cd}\right) \\
\partial_{[\bar b} 
\left(G F_{\ov a] \ov z cd}\tilde{\phi}^{cd}\right)  \\
\end{matrix}
\right)  \\
 \end{array}  \right\} \,,
\end{equation}
where $\tilde{\phi}_{\ov a \ov b}$ denotes harmonic forms that {\it in addition satisfy the constraint} (\ref{eq:harm}). This constraint is the essential difference between the fluxless case and the case with fluxes. As we have mentioned, it presents a linear $h^{2,0}\times h^{2,0}$ system which generically
will have no solution. If we call the dimension of the space of solutions of the constraint equation $n$, the dimension of the space of solutions of the Dirac equation in presence of fluxes becomes $h^{0,0} + h^{1,0} + n + h^{3,0}$.
The undetermined number $n$ obviously satisfies $0\leq n \leq h^{2,0} $. 
The case $n=0$ corresponds to the case in which the flux is completely
generic, the case $n=h^{2,0}$ to the case in which the flux is absent.

\subsection{Counting fermionic zero modes with definite F--chirality}
\label{sec:counting}
Recall that the quantity we wish to determine is $\chi_D(F)$, the sum of the number of solutions to the Dirac equation in the presence of flux weighted by F--chirality.

We first consider the solutions which are independent of the flux. As we saw in the previous subsection, there are $h^{(0,0)}$ such solutions of positive F--chirality and $h^{(1,0)} + h^{(3,0)}$ with negative F--chirality. In addition, there are $n$ solutions of the form 
\ba
\left(
\begin{matrix}
0,
\tilde{\phi}_{\ov a \ov b},
g^{b \ov a} \partial_b 
\left(G F_{\ov a \ov z cd}\tilde{\phi}^{cd}\right) ,
\partial_{[\bar b} 
\left(G F_{\ov a] \ov z cd}\tilde{\phi}^{cd}\right) 
\end{matrix}
\right)^T  \,.
\ea
If we were ignoring the 
transformation of $F$, this would look like a solution with indefinite $U(1)$ charge. 
It involves a part with positive chirality ($\phi^{ab}$) and a part with 
negative chirality (the ones in (\ref{eq:partsol})) (this is of course different from having a solution with $\phi^{ab}$ and one 
with the $\phi^0$, separately). When we take into account the transformation $F \to -F$, we
see that (\ref{eq:partsol}) gives us an extra minus sign, such that all the components have positive F-chirality.

Collecting everything, we arrive at the expression 
\begin{equation}
  \chi_D(F) = h^{0,0}+n(F)-h^{1,0} 
-h^{3,0}
\label{index}
\end{equation}
 for our index, where we have denoted the $F$ dependence of $n$ for emphasis. In the next section, we will study some concrete examples in which we can determine $n$ explicitly.

\section{ Example of $K3\times K3$ compactification }
In this section, we apply the formalism of the previous sections to the specific example of an M5 brane wrapping a 6-cycle $D=K3\times \IP^1$ of the 4-fold $X=K3\times K3$. We introduce the following convention for the various indices in our equations: for the tangent space to the 6-cycle $D$, $a, b; \ov a, \ov b$ $(1,2; \ov 1, \ov 2)$ for directions tangent to $K3$, $z_1, \ov z_1$ $(3; \ov 3)$ for directions tangent to $\IP^1$, and for the normal directions within $X$ to the 6-cycle, $z_2, \ov z_2$ $(4;\ov 4)$. When we need to distinguish between the two $K3$'s, we introduce indices $K3_1 \times K3_2$, and refer e.g. to the complex structures as $\Omega_1$, $\Omega_2$ respectively.

Since both $K3$ and $\IP^1$ only have even cohomology, the same is true for the cohomology of the 6-cycle by the K\"unneth formula. The Dirac equation in the fluxless case hence only has positive chirality solutions, i.e. $\phi_{\ov 1 \ov 3}= \phi_{\ov 2 \ov 3}=0$, and we have
\begin{equation}
  \epsilon_+= \phi |\Omega\rangle + \phi_{\ov 1 \ov 2}\Gamma^{\ov 1 \ov 2}|\Omega\rangle  \,.
\end{equation}
We will consider two cases: fluxes that preserve respectively all or half of the supersymmetry of the background.
\begin{enumerate}
\item{To preserve the full $N=2$ supersymmetry (again, we are using 4d terminology, i.e. $N=2 \rightarrow$ 8 supercharges), we consider flux which is a $(1,1)$ form on $K3$ \cite{Dasgupta:1999ss,Tripathy:2002qw,Angelantonj:2003zx}. 
In a local patch, a possible choice is that all components vanish other than
\begin{equation}
F_{13\ov 2\ov 4}\neq 0 \,,
\label{susynonbreak}
\end{equation}
and its complex conjugate.


According to our analysis in section \ref{sec:general}, we can find $h^{0,0}$ solutions of even F-chirality and $h^{0,1}+h^{0,3}$ solutions of odd F-chirality to the Dirac equation for any choice of flux. To determine whether we have additional solutions, we need to find  harmonic $(0,2)$ forms $\phi_{\ov b \ov c}$ which satisfy
\begin{equation}
  {\cal H} (F_{\ov a \ov z bc}\phi^{bc}dz^a)=0 \;. \label{constr}
\end{equation}
The only harmonic $(0,2)$ form on $D$ is $\ov \Omega_1$.
The contraction of our choice of flux (which is a (1,1) form on the $K3$) with $\ov \Omega_1$ vanishes. Hence, eq (\ref{constr}) is trivially satisfied, and $\ov \Omega_1$ is a further solution to the Dirac equation, with positive F-chirality.

Overall, the number of solutions to the Dirac equation hence does not change upon turning on this choice of flux. In particular, 
$\chi_D = \chi_D(F)=2$.}

\item{To break the $N=2$ supersymmetry preserved by $X$ to $N=1$, we consider a flux which is a $(2,0)+(0,2)$ form on $K3$ \cite{Dasgupta:1999ss,Tripathy:2002qw,Angelantonj:2003zx}, e.g., in a local patch, all components vanish other than
\begin{equation}
F_{12\ov 3\ov 4}\neq 0
\label{susybreak}
\end{equation}
and its complex conjugate.
In particular, we consider the choice
\begin{equation}
  {F\over  \pi} =\Omega_1\wedge \bar \Omega_2 + \bar \Omega_1\wedge \Omega_2
\label{4form}
\end{equation}
made in  \cite{Gorlich:2004qm}.

As before, we need to contract the flux with $\ov \Omega_1$ and project onto the non-harmonic piece.
This amounts to contracting $\Omega_1$ with $\ov \Omega_1$, which is a number by covariant constancy of the complex structure. We are left with $\ov \Omega_2$, the harmonic projection of which is itself. $\ov \Omega_1$ hence does not solve eq (\ref{eq:harm}) and is not a solution of our Dirac equation.

In this example, we hence lose a zero mode of the Dirac operator, namely $\phi_{\ov a \ov b} \Gamma^{\ov a \ov b} | \Omega \rangle$, upon turning on flux. In particular, $\chi_D =2$, while $\chi_D(F) =1$.\footnote{This example is 
simpler than the general case, because the only zero modes present 
are those which come from supersymmetry; that is, they are pullbacks 
of spinors from the bulk. The fact that half of 
these spinors are lifted by the flux is simply the statement 
that half of the supercharges are broken by it. We thank Savdeep Sethi
for sharing this argument with us. In general, when there are additional zero
modes which do not come from supersymmetry, the analysis given in this paper
will be necessary. We hope to report on examples for this general case in the 
 future.}
}
\end{enumerate}

\section{Discussion} 

This work may have  important consequences for  properties of the string theory landscape and for stringy cosmology. Initially, it was motivated by the study of the stabilization of moduli in D3/D7 cosmological model. However, our results have  general validity: we have found  the conditions for the existence of non-perturbative corrections to the superpotential in the general case when fluxes are present. Our only restriction to fluxes in M-theory compactified on a CY 4-fold which are  (2,2) and primitive, as required for unbroken $N=1$ supersymmetry \cite{Becker:1996gj}.

Our result for the counting of the fermionic zero modes (weighted with the U(1) charge) differs from the result in the absence of fluxes \cite{Witten:1996bn}  where it is given by $\chi_D(F)= \chi_{D}$. We find
\begin{equation}
  \chi_D(F)= \chi_{D}-(h^{(0,2)}-n)
\label{number}
\end{equation}
Here $n$ is the dimension of solutions of the eq. (\ref{eq:harm})  which depends on fluxes. In absence of fluxes  $n=h^{(0,2)}$ and we reproduce the $\chi_D(F)= \chi_{D}$ formula.

In our example of $K3\times K3$ compactification  we have found cases with $n=0$ and $\chi_D(F)=h^{(0,0)}+h^{(0,2)}-h^{(0,2)}=1$ when fluxes break $N=2$ supersymmetry of the manifold down to $N=1$. We have also found cases with $n=h^{(0,2)}=1$ and $\chi_D(F)=2$, when fluxes preserve $N=2$ supersymmetry of the background. In the first case we see that instanton corrections are possible for a divisor on $K3\times \IP^1$. This indicates that in the IIB formulation the model with $K3\times {T^2\over \IZ^2}$ compactification will have instanton corrections for K\"ahler moduli: in the M-theory they would correspond to $K3_1\times \IP^1$ as well as $\IP^1 \times K3_2$ divisors. We hope to confirm this conclusion  by also generalizing our methods to the counting of the fermionic zero modes on D3 branes with fluxes.

Our answer in the general case suggests that one may try to look for instanton corrections by requiring that
\begin{equation}
  \chi_D(F)= \chi_{D}-(h^{(0,2)}-n)=1\ .
\label{number1}
\end{equation}
This is reduced to 
\begin{equation}
 n= h^{(0,1)}+ h^{(0,3)} \ .
\label{inst}
\end{equation}

The existence of the new class of string vacua with $\chi_D\not =1$, $\chi_D(F) =1$ may further enhance the richness of the stringy landscape and may help us to find a full description of the  realistic models of string inflation, including the D3/D7 model.

\

{\it Note added}:  When this paper was at the stage of completion, the paper \cite{Saulina:2005ve} appeared with some overlap with our results.

\

\leftline{\bf Acknowledgements} We are grateful to P. Aspinwall, E. Bergshoeff, S. Ferrara, S. Gukov, S. Kachru, A. Linde, J. McGreevy,
G. Moore,  S. Sethi, I.~Singer, and S. Trivedi  for
useful discussions. This work was supported by NSF
grant 0244728. The work of AK was also supported by the U.S. Department of Energy under contract number DE-AC02-76SF00515.


\begin{thebibliography}{999}

\bibitem{Giddings:2001yu}
  S.~B.~Giddings, S.~Kachru and J.~Polchinski,
``Hierarchies from fluxes in string compactifications,''
  Phys.\ Rev.\ D {\bf 66}, 106006 (2002)
  [arXiv:hep-th/0105097].

\bibitem{Kachru:2004jr}
  S.~Kachru and A.~K.~Kashani-Poor,
  ``Moduli potentials in type IIA compactifications with RR and NS flux,''
  arXiv:hep-th/0411279.


\bibitem{Kachru:2003aw}
S.~Kachru, R.~Kallosh, A.~Linde and S.~P.~Trivedi,
``De Sitter vacua in string theory,''
Phys.\ Rev.\ D {\bf 68}, 046005 (2003)
[arXiv:hep-th/0301240].

\bibitem{Denef:2005mm}
  F.~Denef, M.~R.~Douglas, B.~Florea, A.~Grassi and S.~Kachru,
``Fixing All Moduli in a Simple F-Theory Compactification,''
  arXiv:hep-th/0503124.




\bibitem{Herdeiro:2001zb}
  C.~Herdeiro, S.~Hirano and R.~Kallosh,
``String theory and hybrid inflation / acceleration,''
  JHEP {\bf 0112}, 027 (2001)
  [arXiv:hep-th/0110271].
  K.~Dasgupta, C.~Herdeiro, S.~Hirano and R.~Kallosh,
``D3/D7 inflationary model and M-theory,''
  Phys.\ Rev.\ D {\bf 65}, 126002 (2002)
  [arXiv:hep-th/0203019].
  J.~P.~Hsu, R.~Kallosh and S.~Prokushkin,
``On brane inflation with volume stabilization,''
  JCAP {\bf 0312}, 009 (2003)
  [arXiv:hep-th/0311077].
H.~Firouzjahi and S.~H.~H.~Tye, ``Closer towards inflation in string theory,''
[arXiv:hep-th/0312020].
  J.~P.~Hsu and R.~Kallosh,
``Volume stabilization and the origin of the inflaton shift symmetry in
string theory,''
  JHEP {\bf 0404}, 042 (2004)
  [arXiv:hep-th/0402047].
  K.~Dasgupta, J.~P.~Hsu, R.~Kallosh, A.~Linde and M.~Zagermann,
``D3/D7 brane inflation and semilocal strings,''
  JHEP {\bf 0408}, 030 (2004)
  [arXiv:hep-th/0405247].

\bibitem{Angelantonj:2003zx}
C.~Angelantonj, R.~D'Auria, S.~Ferrara and M.~Trigiante,
 ``K3 x T**2/Z(2) orientifolds with fluxes, open string moduli  and critical
points,'' [arXiv:hep-th/0312019];
R.~D'Auria, S.~Ferrara and M.~Trigiante,
``Homogeneous special manifolds, orientifolds and solvable coordinates,''
arXiv:hep-th/0403204.
  R.~D'Auria, S.~Ferrara and M.~Trigiante,
  ``No-scale supergravity from higher dimensions,''
  arXiv:hep-th/0409184.

\bibitem{Koyama:2003yc}
  F.~Koyama, Y.~Tachikawa and T.~Watari,
  ``Supergravity analysis of hybrid inflation model from D3-D7 system,''
  Phys.\ Rev.\ D {\bf 69}, 106001 (2004)
  [Erratum-ibid.\ D {\bf 70}, 129907 (2004)]
  [arXiv:hep-th/0311191].



\bibitem{Witten:1996bn}
E.~Witten,
``Non-Perturbative Superpotentials In String Theory,''
Nucl.\ Phys.\ B {\bf 474}, 343 (1996)
[arXiv:hep-th/9604030].

\bibitem{Gorlich:2004qm}
  L.~Gorlich, S.~Kachru, P.~K.~Tripathy and S.~P.~Trivedi,
``Gaugino condensation and nonperturbative superpotentials in flux
compactifications,''
  arXiv:hep-th/0407130.

\bibitem{Robbins:2004hx}
  D.~Robbins and S.~Sethi,
  ``A barren landscape,''
  Phys.\ Rev.\ D {\bf 71}, 046008 (2005)
  [arXiv:hep-th/0405011].


\bibitem{Kallosh:2005yu}
  R.~Kallosh and D.~Sorokin,
 ``Dirac action on M5 and M2 branes with bulk fluxes,''
  arXiv:hep-th/0501081.

\bibitem{Tripathy:2005hv}
  P.~K.~Tripathy and S.~P.~Trivedi,
  ``D3 brane action and fermion zero modes in presence of background flux,''
  arXiv:hep-th/0503072.

\bibitem{Sorokin:1999jx}
D.~P.~Sorokin,
``Superbranes and superembeddings,''
Phys.\ Rept.\  {\bf 329}, 1 (2000)
[arXiv:hep-th/9906142].




\bibitem{Becker:1996gj}
  K.~Becker and M.~Becker,
  ``M-Theory on Eight-Manifolds,''
  Nucl.\ Phys.\ B {\bf 477}, 155 (1996)
  [arXiv:hep-th/9605053].

\bibitem{Dasgupta:1999ss}
  K.~Dasgupta, G.~Rajesh and S.~Sethi,
  ``M theory, orientifolds and G-flux,''
  JHEP {\bf 9908}, 023 (1999)
  [arXiv:hep-th/9908088].



\bibitem{Tripathy:2002qw}
  P.~K.~Tripathy and S.~P.~Trivedi,
  ``Compactification with flux on K3 and tori,''
  JHEP {\bf 0303}, 028 (2003)
  [arXiv:hep-th/0301139].


\bibitem{Saulina:2005ve}
  N.~Saulina,
  ``Topological constraints on stabilized flux vacua,''
  arXiv:hep-th/0503125.



\end{thebibliography}
\end{document}